\input{aipcheck}
\documentclass[final,mathptmx,4apaper]{aipproc}

\layoutstyle{6x9}

\begin{document}

\title{Multiparticle Quantum Cosmology\footnote{Given as a talk "Many Particle Quantum Cosmology" at the Ninth International Symposium \emph{Frontiers of Fundamental and Computational Physics}, Udine University, Udine, Italy , and The Abdus Salam International Center for Theoretical Physics, Trieste, Italy, January 7-9, 2008}}


\author{L. A. Glinka\footnote{E-mail to: laglinka@gmail.com}}{address={N.N. Bogoliubov Laboratory of Theoretical Physics, Joint Institute for Nuclear Research,\\Joliot--Curie 6, 141980 Dubna, Moscow Region, Russian Federation}, email={laglinka@gmail.com}, thanks={Talk \emph{Many particle quantum cosmology} at the Ninth International Symposium \emph{Frontiers of Fundamental and Computational Physics}, January 7-9, 2008, Udine University in Udine, and The Abdus Salam ICTP in Trieste, Italy}}

\begin{abstract}
Fock space quantization of Hamiltonian constraints of General Relativity and thermodynamics of quantum states for flat Friedmann--Lema\^{i}tre--Robertson--Walker metrics is presented.
\end{abstract}

\maketitle


\section{Introduction. Thermodynamical Einstein's Dream}
General Relativity is a theory of some Riemannian manifold \cite{rie} with dynamics governs by the Einstein equations \cite{ein1} thats can be deduced from the Hilbert dynamical action \cite{hil} by the Palatini variational principle \cite{mtw}
\begin{eqnarray}
\mathit{S}_{\mathrm{EH}}&=&\int d^4x\sqrt{-\det g_{\mu\nu}}\left(-\frac{1}{6}R+\mathcal{L}\right),\label{eh}\\
\frac{\delta S_{\mathrm{EH}}}{\delta g^{\mu\nu}}&=&0\Longrightarrow R_{\mu\nu}-\frac{1}{2}g_{\mu\nu}R=3T_{\mu\nu},~~T_{\mu\nu}=\frac{2}{\sqrt{-\det g_{\mu\nu}}}\frac{\delta\left( \sqrt{-\det g_{\mu\nu}}\mathcal{L}\right)}{\delta g^{\mu\nu}},\label{eh1}
\end{eqnarray}
where $R_{\mu\nu}$ is the Ricci tensor, $R=g^{\alpha\beta}R_{\alpha\beta}$ is the Ricci curvature scalar, $\mathcal{L}$ is a full Lagrangian of all physical fields, and $T_{\mu\nu}$ is the stress--energy tensor of these fields \footnote{In this paper the units $8\pi G/3=c=\hbar=k_B=1$ are used.}.

Let us to take up address an issue of a Riemannian manifold given by the flat Friedmann--Lema\^{i}tre--Robertson--Walker metric \cite{frw} characterized by an interval\footnote{For closed case $V_0=\int d^3x<\infty$ the metrics (\ref{ef}) describe \emph{cylindrical Einstein--Friedmann Universe.} \cite{ein2}}
\begin{equation}\label{ef}
ds^2=g_{\mu\nu}dx^{\mu}dx^{\nu}=(dx^0)^2-a^2(t)(dx^i)^2,
\end{equation}
where $a$ is the Friedmann conformal scale factor and $x^\mu$, $\mu=0,1,2,3$ is a cartesian coordinate system. In the canonical Dirac--ADM approach to General Relativity \cite{adm} the metric (\ref{ef}) is characterized by vanishing shift function and lapse function $N$ is given by
\begin{equation}\label{ct}
d\eta=N(x^0)dx^0\equiv\frac{dt}{a(t)},~~\mathrm{where}~~t=\tau+x^0,~~\tau=\mathit{const},
\end{equation}
where $t$ is called cosmological time, and $\eta$ is called conformal time. The metrics (\ref{ef}) in the Dirac-ADM approach are equivalent to the following Hamiltonian constraints
\begin{equation}\label{c}
\mathrm{H}={p}_{a}^{2}-4V_0^2a^4H^2(a)=0,
\end{equation}
where ${p}_a$ is canonical momentum conjugated to $a$, and $H(a)$ is the Hubble parameter,
\begin{equation}
{p}_{a}=-2V_0\frac{da}{d\eta},~~H^2(a)\equiv\left(\frac{1}{a}\frac{da}{dt}\right)^2=\left(\frac{1}{a^2}\frac{da}{d\eta}\right)^2=\frac{1}{V_0}\int{d^{3}x}~\mathcal{H}(x)
\end{equation}
where $\mathcal{H}(x)$ is summarized Hamiltonian of physical fields associated by the Legendre transformation with the Lagrangian $\mathcal{L}$ in the dynamical action (\ref{eh}). The Hubble law in the canonical approach is a result of integration of (\ref{c}) and has a form
\begin{equation}\label{hub}
\int_{a_I}^{a}\frac{da^\prime}{a^\prime H(a^\prime)}=t_I-t,~~\mathrm{where}~I~\mathrm{means~initial~data}.
\end{equation}

We propose call by the \emph{Thermodynamical Einstein's Dream (TED)} a some formal road from General Relativity (\ref{eh}-\ref{eh1}) to generalized thermodynamics of quantum states of a Riemannian manifold being a solution of this theory. The questions arise
\begin{itemize}\item \emph{Could TED really be realize?}
\item \emph{Can be exist a way that allows to get to know thermal properties of Spacetime?}
\item \emph{If these properties of Spacetime give new opportunities for Quantum Gravity?}
\end{itemize}
In this paper a proposal for realization of TED is demonstrated by using of second quantization in the Fock space and associated with this quantization formalism of the Bose condensation, where a \emph{particle} is quantum state of a Spacetime, is discussed.
\section{Constraints, strings, and second quantization}
Let us consider the Dirac--ADM primary constraints (\ref{c}). Applying to this equation the formal identification
\begin{equation}
  -4V_0^2a^4H^2(a)=m^2(a)
\end{equation}
allows to equate the geometrical object (\ref{ef}) with free bosonic string with negative square of mass, called a \emph{tachyon} \cite{str}, defined by the constraints
\begin{equation}\label{pc}
p_a^2+m^2=E^2,~~E=0,~~m^2=m^2(a)\leq0.
\end{equation}
The Hubble law (\ref{hub}) can be treated for tachyon as definition of spatial volume $V_0$
\begin{equation}
\left(\int_{a_I^2}^{a^2}\frac{dy}{\left|m(y)\right|}\right)^{-1}\Delta t=V_0,~~\Delta t=t-t_I.
\end{equation}
First quantization of the constraints (\ref{pc}) gives the Wheeler--DeWitt equation \cite{wdw} for (\ref{ef})
\begin{equation}\label{wdw}
i\left[\mathrm{p}_{a},a\right]=1~~\Longrightarrow~~\left[-\frac{\partial^2}{\partial a^2}+m^2(a)\right]\Psi=0,~~\Psi=\Psi^{\ast},
\end{equation}
that can be treated as the free-time Schr\"odinger equation \cite{hh}. However, this point of view has no physical sense. Equally well the Wheeler--DeWitt equation (\ref{wdw}) can be understood as the $0+1$ Klein--Gordon--Fock equation, that treats a Riemannian manifold as a Bose field. Separation of variables method applied to the equation (\ref{wdw}) gives
\begin{equation}\label{dir}
 \left(i\Gamma^{\mu}\partial_{\nu}-\mathrm{M}^{\mu}_{\nu}\right)\Phi_{\mu}=0,~~\mathrm{M}^{\mu}_{\nu}=\left[\begin{array}{cc}
0&-1\\-m^{2}&0\end{array}\right]\geq0
\end{equation}
where matrices $\Gamma^{\mu}=\left[\mathbf{0}_{2},i\mathbf{I}_2\right]$ create Clifford algebra $\left\{\Gamma^{\mu},\Gamma^{\nu}\right\}=2\eta^{\mu\nu}\mathbf{I}_2$, and was introduced the state vector $\Phi_{\mu}=\left[\Psi,~\Pi_\Psi\right]^{T}$ with $\Pi_{\Psi}$ being classical canonical momentum field conjugated to wave function $\Psi$, here also $\partial_{\nu}=\left[0,-\frac{\partial}{\partial a}\right]^{T}$. Let us carry second quantization \cite{bos}  of the equation (\ref{dir}) that is agreed with (\ref{wdw}). This has a proposed form
\begin{eqnarray}\label{sq}
\left[\begin{array}{c}\hat{\Psi}[a]\\\hat{\Pi}_\Psi[a]\end{array}\right]
=\left[\begin{array}{cc}1/\sqrt{2|m(a)|}&1/\sqrt{2|m(a)|}\\
-i\sqrt{|m(a)|/2}&i\sqrt{|m(a)|/2}\end{array}\right]
\left[\begin{array}{c}{G}[a]\\ {G}^{\dagger}[a]\end{array}\right],
\end{eqnarray}
and automatically fulfills principal rule of quantum field theory of the Bose systems
\begin{equation}
  \left[\hat{\Pi}_{\Psi}[a],\hat{\Psi}[a']\right]=-i\delta\left(a-a'\right),~~a\equiv a(\eta),~~a^\prime\equiv a(\eta^\prime)
\end{equation}
and lies in accordance with general algebraic approach investigated in papers of Von Neumann, Araki and Woods \cite{ccr}. In result, considered Bose system is described by the dynamical basis in the Fock space
\begin{equation}\label{db}
  {B}_{a}=\left\{\left[\begin{array}{c}{G}[a]\\
{G}^{\dagger}[a]\end{array}\right]:\left[{G}[a],{G}^{\dagger}[a']\right]=\delta\left(a-a'\right), \left[{G}[a],{G}[a']\right]=0\right\},
\end{equation}
according to the evolution equation
\begin{equation}\label{df}
\frac{\partial {B}_{a}}{\partial{a}}=\left[\begin{array}{cc}
-im&\frac{1}{2m}\frac{\partial m}{\partial a}\\
\frac{1}{2m}\frac{\partial m}{\partial a}&im\end{array}\right]{B}_{a},
\end{equation}
that can be diagonalized to the Heisenberg form by the Bogoliubov transformation
\begin{equation}
{B}_{a}^\prime=\left[\begin{array}{cc}u&v\\
v^{\ast}&u^{\ast}\end{array}\right]{B}_{a},~~\frac{\partial{B}_{a}^\prime}{\partial a}=\left[\begin{array}{cc}
-i\lambda & 0 \\ 0 &
i\lambda\end{array}\right]{B}_{a}^\prime,
\end{equation}
where $|u|^2-|v|^2=1$, and
\begin{equation}
  {B}_{a}^\prime=\left\{\left[\begin{array}{c}{G}^\prime[a]\\
{G}^{\prime\dagger}[a]\end{array}\right]:\left[{G}^\prime[a],{G}^{\prime\dagger}[a']\right]=\delta\left(a-a'\right), \left[{G}^\prime[a],{G}^\prime[a']\right]=0\right\},
\end{equation}
is some new basis, and the Bogoliubov coefficients $u$ and $v$ are functions of $a$. This procedure gives simply $\lambda=0$, and by this ${G}^{\prime}[a]=\mathrm{w}_I=const$ defines stable vacuum $|0\rangle_I=\left\{|0\rangle_I:\mathrm{w}_I|0\rangle_I=0,~0={_I}\langle0| \mathrm{w}_I^\dagger\right\}$ in the static Fock basis ${B}_a^\prime={B}_{I}$,
\begin{equation}\label{in}
{B}_{I}=\left\{\left[\begin{array}{c}\mathrm{w}_I\\
\mathrm{w}^{\dagger}_I\end{array}\right]: \left[\mathrm{w}_I,\mathrm{w}^{\dagger}_I\right]=1, \left[\mathrm{w}_I,\mathrm{w}_I\right]=0\right\}.
 \end{equation}
As the result the system of equations for the Bogoliubov coefficients is obtained in the form
\begin{equation}\label{bcof}
  \frac{\partial}{\partial a}\left[\begin{array}{c}u\\v\end{array}\right]=\left[\begin{array}{cc}
-im&\frac{1}{2m}\frac{\partial m}{\partial a}\\
\frac{1}{2m}\frac{\partial m}{\partial a}&im\end{array}\right]\left[\begin{array}{c}u\\v\end{array}\right],
\end{equation}
and by applying of the hyperbolic parametrization, this system can be solved directly as
\begin{eqnarray}\label{sup}
u(a)&=&\frac{1}{2}\exp\left\{\pm i\int_{a_I}^{a}mda\right\}\left(\sqrt{\left|\frac{m}{m_I}\right|}+\sqrt{\biggr|\frac{m_I}{m}\biggr|}\right),\\
v(a)&=&\frac{1}{2}\exp\left\{\pm i\int_{a_I}^{a}mda\right\}\left(\sqrt{\left|\frac{m}{m_I}\right|}-\sqrt{\biggr|\frac{m_I}{m}\biggr|}\right),
\end{eqnarray}
where $m=m(a)$, and $m_I=m(a_I)$. In this manner, in proposed approach \emph{quantum gravity is defined by monodromy between bases in the Fock space}
\begin{eqnarray}\label{mon}
{B}_a\!=\!\left[\!\!\!\!\begin{array}{cc}
\left(\sqrt{\left|\frac{m_I}{m(a)}\right|}+\sqrt{\left|\frac{m(a)}{m_I}\right|}\right)\frac{e^{-i\lambda(a)}}{2}\vspace*{5pt}\!\!\!&
\left(\sqrt{\left|\frac{m_I}{m(a)}\right|}-\sqrt{\left|\frac{m(a)}{m_I}\right|}\right)\frac{e^{i\lambda(a)}}{2}\!\!\\
\left(\sqrt{\left|\frac{m_I}{m(a)}\right|}-\sqrt{\left|\frac{m(a)}{m_I}\right|}\right)\frac{e^{-i\lambda(a)}}{2}\!\!\!&
\left(\sqrt{\left|\frac{m_I}{m(a)}\right|}+\sqrt{\left|\frac{m(a)}{m_I}\right|}\right)\frac{e^{i\lambda(a)}}{2}\end{array}\!\!\right]\!\!{B}_I.
\end{eqnarray}
where for compact notation $\lambda(a)=\pm\int_{a_I}^{a}m(a^\prime)~da^\prime$. The field operator $\hat{\Psi}$ that represents the Spacetime (\ref{ef}) as boson and defines quantum states of the manifold is
\begin{eqnarray}
  \hat{\Psi}[a]=\frac{1}{2m(a)}\sqrt{\frac{m_I}{2}}\left(e^{-i\lambda(a)}\mathrm{w}_I+e^{i\lambda(a)}\mathrm{w}^{\dagger}_I\right).
\end{eqnarray}
\section{One-particle thermodynamics of quantum states}
Initial data given by the basis (\ref{in}) define thermal equilibrium state for ensemble of quantum states of the manifold (\ref{ef}). Let us consider thermal properties of this Spacetime - the final step of the Thermodynamical Einstein's Dream. Here we will study one-particle approximation \cite{bos} defined by density functional $\varrho_{{G}}$ in initial data basis
\begin{eqnarray}
\varrho_{{G}}={G}^{\dagger}[a]{G}[a]={B}_{a}^{\dagger}\left[\begin{array}{cc}1&0\\0&0\end{array}\right]{B}_{a}
={B}_I^{\dagger}\left[\begin{array}{cc}|u|^2&-uv\\-u^{\ast}v^{\ast}&|v|^2\end{array}\right]{B}_I\equiv{B}_I^{\dagger}\rho_{\mathrm{eq}}{B}_I.
\end{eqnarray}
Equilibrium entropy $\mathrm{S}$ and partition function $\Omega_{\mathrm{eq}}$ of the system in initial data basis are
\begin{equation}
  \mathrm{S}=-\frac{\mathrm{tr}\left(\rho_{\mathrm{eq}}\ln\rho_{\mathrm{eq}}\right)}{\mathrm{tr}\rho_{\mathrm{eq}}}\equiv\ln\Omega_{\mathrm{eq}},~~\Omega_{\mathrm{eq}}=\frac{1}{2|u|^2-1}.
\end{equation}
Physical quantum states of considered manifold are assumed to be described in the Gibbs ensemble. One can compute directly the entropy of the system
\begin{equation}\label{ent}
  \mathrm{S}=-\ln\langle\mathrm{n}\rangle,~~\langle\mathrm{n}\rangle=2\mathrm{n}+1.
\end{equation}
Here $\langle\mathrm{n}\rangle$ is averaged  occupation number $\mathrm{n}$ of quantum states determined by
\begin{equation}
\mathrm{n}\equiv\langle0|{G}^{\dagger}[a]{G}[a]|0\rangle=\frac{1}{4}\left(\left|\frac{m}{m_I}\right|+\bigg|\frac{m_I}{m}\bigg|\right)-\frac{1}{2},\label{s}
\end{equation}
that together with natural conditions $n\geq0$ and $|m|\geq|m_I|$ leads to the \emph{energy spectrum}
\begin{equation}
  \left|\frac{m}{m_I}\right|=\left|\frac{mc^2}{m_Ic^2}\right|=\langle\mathrm{n}\rangle+\sqrt{\langle\mathrm{n}\rangle^2-1},
\end{equation}
and internal energy $\mathrm{U}$ and chemical potential $\mu$ computed by proper averaging are
\begin{eqnarray}\label{n}
\mathrm{U}&=&\left[1+\left(2+\frac{1}{\langle\mathrm{n}\rangle}\right)(\langle\mathrm{n}\rangle-1)\right]\left(\langle\mathrm{n}\rangle+\sqrt{\langle\mathrm{n}\rangle^2-1}\right)\frac{|m_I|}{2},\label{u}\\
\mu&=&\left[1+\frac{1}{\langle\mathrm{n}\rangle^2}-\left(2-\frac{1}{\langle\mathrm{n}\rangle}\right)\sqrt{\frac{1}{\langle\mathrm{n}\rangle^2-1}}\right]\left(\langle\mathrm{n}\rangle+\sqrt{\langle\mathrm{n}\rangle^2-1}\right)|m_I|.
\end{eqnarray}
From bosonic property one can assume in equilibrium the Bose--Einstein statistics
\begin{equation}
\Omega_{\mathrm{eq}}\equiv\left(\exp\left\{\frac{\mathrm{U}-\mu\mathrm{n}}{\mathrm{T}}\right\}-1\right)^{-1},
\end{equation}
and by this one can compute directly temperature of the system
\begin{equation}
  \mathrm{T}=\frac{\langle\mathrm{n}\rangle+\sqrt{\langle\mathrm{n}\rangle^2-1}}{\ln(\langle\mathrm{n}\rangle+1)}\left[1+\left(1-\frac{1}{\langle\mathrm{n}\rangle}\right)^2+\left(\langle\mathrm{n}\rangle-\frac{1}{2\langle\mathrm{n}\rangle}\right)\sqrt{\frac{\langle\mathrm{n}\rangle-1}{\langle\mathrm{n}\rangle+1}}\right]\frac{|m_I|}{2}.\label{t}
\end{equation}

\emph{Maximal entropy quantum states (MEQS) of the Riemannian manifold} are defined by
\begin{equation}\label{max}
  \mathrm{S}_{\mathrm{max}}=0 \Longrightarrow n=0,~~ \langle\mathrm{n}\rangle=1,~~m=m_I,~~\mathrm{U}_{\mathrm{max}}=\frac{|m_I|}{2},~~\mu_{\mathrm{max}}=-\infty,~~\mathrm{T}_{\mathrm{max}}=\frac{\mathrm{U}_{\mathrm{max}}}{\ln2}.
\end{equation}
These allow conclude that \emph{for the metrics MEQS are fully determinated by initial data.} The last equation in (\ref{max}) is an equation of state for MEQS, that describes classical ideal Bose gas. Infinite value of chemical potential means that MEQS create closed system, and the point $\mathrm{n}=0$ is phase transition point for MEQS set. Initial data point can be interpreted as a birth-point of the Riemannian manifold. Measurement of MEQS formally defines initial data point, and it defines initial data reference frame. Furthermore, MEQS set have very interesting physical interpretation that is a conclusion of the reasoning
\begin{equation}\label{maxi}
  m_I=m=i2V_0a^2H(a) \Longrightarrow H(a)=\frac{Q}{a^2},~~Q=\frac{|m_I|}{2V_0}=\frac{\mathrm{U}_{\mathrm{max}}}{V_0}=a_I^2H(a_I),
\end{equation}
and for $\mathrm{U}_{\mathrm{max}}=\mathrm{p}_{\mathrm{max}}V_0$ it gives pressure $\mathrm{p}_{\mathrm{max}}=a_I^2H(a_I)$. Form of the Hubble parameter $H(a)$ in (\ref{maxi}) means that \emph{set of MEQS for the manifold describes primordial radiation.} Initial data basis ${B}_I$ is related with this light. In light of modern observational cosmology one can identify this primordial light with CMB radiation. One can see that $\lambda$ in (\ref{mon}) is linear for MEQS and fulfills the d'Alembert equation
\begin{equation}
\lambda(a)=\left.\int_{a_I}^a m(a^\prime)da^\prime\right|_{m=m_I}=m_I(a-a_I)\Rightarrow\Delta\lambda=0,~~a(t)=a_I\sqrt{1+2Q(t-t_I)}
\end{equation}
and by this can be treated as the Weyl scale \cite{pau} of the ensemble.

Quantum Gravity defined by the second quantization (\ref{sq}) used to the metrics (\ref{ef}) gives beautiful scenario for physics of the Universe. It allows to state that TED is realizable for considered manifold. From the String Theory point of view \emph{tachyon is mass groundstate of bosonic string that describes some wider quantum theory of gravity}. Basing on this let one create the following conjecture: \emph{for the Einstein--Hilbert (\ref{eh1}) gravitational fields analogical deduction can be applied}. Moreover, multiparticle reasoning creates understanding of gravity by quantum field theory, and describes General Relativity by the Bose condensates. This approach can be called Multiparticle Quantum Gravity/Cosmology.

\begin{theacknowledgments}
  The author benefitted from valuable discussions with \mbox{B.G. Sidharth}, \mbox{A. De Angelis}, \mbox{J.G. Hartnett}, and \mbox{J. Lewandowski} and is grateful to \mbox{B.G. Sidharth} for invitation to the 9th International Symposium \emph{Frontiers of Fundamental and Computational Physics}.
\end{theacknowledgments}

\bibliographystyle{aipproc}   

\begin{thebibliography}{99}
\bibitem{rie}   B. Riemann, \textit{Nachr. Ges. Wiss. G\"ottingen} \textbf{13}, 133, (1920).

\bibitem{ein1}  A. Einstein, \textit{Sitzungsber. Preuss. Akad. Wiss. Berlin} \textbf{44}, N2, 778, (1915); \textit{Sitzungsber. Preuss. Akad. Wiss. Berlin} \textbf{46}, N2, 799, (1915); \textit{Sitzungsber. Preuss. Akad. Wiss. Berlin} \textbf{48}, N2, 844, (1915); \textit{Ann. Phys.} \textbf{49}, 769, (1916); \textit{Rev. Mod. Phys.} \textbf{20}, 35, (1948).

\bibitem{hil}   D. Hilbert, \textit{Konigl. Gesell. d. Wiss. G\"ottinger, Nachr., Math.-Phys. Kl.} \textbf{27}, 395, (1915).

\bibitem{mtw}   A. Palatini, \textit{Rend. Pal.} \textbf{43}, 203, (1919).

\bibitem{frw}   A. Friedmann, \textit{Z. Phys.} \textbf{10}, 377, (1922); \textit{Z. Phys.} \textbf{21}, 326, (1924); G. Lema\^{i}tre, \textit{J. Math. Phys} \textbf{4}, 188, (1925); \textit{Phys. Rev.}, \textbf{25}, 903 (1925); \textit{Ann. Soc. Sci. Brux. A}, \textbf{47}, 49, (1931); \textit{Ann. Soc. Sci. Brux. A} \textbf{53}, 51, (1933); H.P. Robertson, \textit{Phil. Mag.}, \textbf{5}, 835, (1928); \textit{Astrophys. Jour.} \textbf{82}, 248, (1935); \textit{Astrophys. Jour.} \textbf{83}, 187, (1936); \textit{Astrophys. Jour.} \textbf{83}, 257, (1936); A.G. Walker, \textit{Proc. Lon. Math. Soc. 2}, \textbf{42}, 90, (1937).

\bibitem{adm}   P.A.M. Dirac, \textit{Proc. Roy. Soc. Lond. A} \textbf{246}, 333, (1958); \textit{Phys. Rev.} \textbf{114}, 924, (1959); \textit{Proc. Roy. Soc. Lond. A} \textbf{246}, 326, (1958); \textit{Can. J. Math.} \textbf{2}, 129, (1950); R. Arnowitt, S. Deser and C.W. Misner, in \textit{Gravitation: an introduction to current research}, edited by L. Witten (John Wiley and Sons, New York, 1961), p. 227

\bibitem{ein2}  A. Einstein, The meaning of relativity, Princeton University Press, Princeton (1922).
\bibitem{bos}   J.-P. Blaizot and G. Ripka, \textit{Quantum theory of f\/inite systems}, Massachusetts Institute of Technology Press, (1986).

\bibitem{str}   D. L\"ust and S. Theisen, \textit{Lect. Notes Phys.} \textbf{346}, (1989).

\bibitem{wdw}   J.A. Wheeler, in \textit{Battelle Rencontres: 1967  Lectures in Mathematics and Physics}, Editors C.M. DeWitt and J.A. Wheeler, New York, 1968, p. 242; B.S. DeWitt, \textit{Phys. Rev.} \textbf{160}, 1113, (1967).

\bibitem{hh}    A. Ashtekar, M. Bojowald and J. Lewandowski, \textit{Adv. Theor. Math. Phys.} \textbf{7}, 233, (2003); J. Lewandowski \emph{Private communication}; C. Rovelli \emph{Private communication}

\bibitem{ccr}   J. von Neumann, \textit{Math. Ann.} \textbf{104}, 570, (1931); H. Araki and E.J. Woods, \textit{J. Math. Phys.} \textbf{4}, 637, (1963); J. Derezi\'nski, \textit{Lect. Notes Phys.} \textbf{695}, 63-143, (2005).

\bibitem{pau} Weyl H., \textit{Raum-Zeit-Materie}, 6th ed., Springer-Verlag Berlin-Heidelberg, (1970).

\end{thebibliography}

\IfFileExists{\jobname.bbl}{}
 {\typeout{}
  \typeout{******************************************}
  \typeout{** Please run "bibtex \jobname" to optain}
  \typeout{** the bibliography and then re-run LaTeX}
  \typeout{** twice to fix the references!}
  \typeout{******************************************}
  \typeout{}
 }
 \end{document}